\documentclass[twocolumn,showpacs,prb,footinbib,a4paper,superscriptaddress,floatfix]{revtex4}

\usepackage{amssymb,amsmath}
\usepackage{graphics}
\usepackage{dcolumn}
\usepackage{bm}
\usepackage{epsfig}

\usepackage{color}

\newcommand{\HN}[1]{{#1}}

\begin{document}

\title{Nonequilibrium density matrix in quantum open systems: generalisation
for simultaneous heat and charge steady-state transport}

\author{H. Ness}
\email{herve.ness@kcl.ac.uk}
\affiliation{Department of Physics, Faculty of Natural and Mathematical Sciences,
King's College London, Strand, London WC2R 2LS, UK}
\altaffiliation{European Theoretical Spectroscopy Facility (ETSF), www.etsf.eu}

\begin{abstract}
We suggest a generalisation of the expression of the nonequilibrium density matrix obtained 
by Hershfield's method for the cases where both heat and charge steady state currents
are present in a quantum open system. 
The finite-size quantum system, connected to two temperature and particle reservoirs, 
is driven out of equilibrium by the presence of both a temperature gradient and a chemical 
potential gradient between the two reservoirs. 
We show that the NE density matrix is given by a generalised Gibbs-like ensemble, and is in
full agreement with the general results of the McLennan-Zubarev nonequilibrium ensembles.
The extra non-equilibrium terms are related to the entropy production in the system and 
characterise the fluxes of heat and particle.
An explicit example, for the lowest order expansion, is provide for a model system of non-interacting
fermions. 
\end{abstract}


\maketitle

\section{Introduction}
\label{sec:intro}

When a finite-size quantum system is put into contact with different 
macroscopic temperature and/or particle reservoirs 
(each at their own equilibrium), the system will reach a non-equilibrium 
(NE) time-independent steady state after some time (which is much 
longer than some typical relaxation times of the finite system).
The steady state is obtained from the balance between irreversible 
processes and the driving forces induced by the macroscopic reservoirs. 

The understanding of such irreversible phenomena and of the corresponding NE 
steady state is a long-standing problem in statistical mechanics.
The NE steady state can be seen as its equilibrium counterpart
for different external constraints, in the sense that
an equilibrium state represents a stationary state of a closed system, while
the NE steady state is the time-invariant state of an open system.
This is why the construction of Gibbs-like ensembles for the NE steady 
state has been explored by many authors.
Early attempts, going beyond linear response \cite{Kubo:1966}, 
have been performed by McLennan \cite{McLennan:1959} for 
classical systems and by Zubarev 
\cite{Zubarev:1974,Zubarev:1994,Zubarev:1996,Zubarev:1997,Morozov:1998}
for both classic and quantum systems.
In such approaches, the Gibbsian statistical mechanics method is extended 
to include steady-state NE conditions in the density matrix
leading to the so-called NE statistical operator method (NESOM).
More rigorous analysis of the existence and stability of the NE steady state
have been performed using $C^*$ algebraic 
methods \cite{Ruelle:2000,Tasaki:2003,Frohlich:2003,Tasaki:2006,Maes:2010,Tasaki:2011,Moldoveanu:2011,Cornean:2014,Note:Cornean:2011}.
The existence of conducting steady states has also been critically discussed
in Refs.~[\onlinecite{Kurth:2010,Khosravi:2012}] by using different levels of 
approximation for the many-body effects in NE Green's functions approaches 
and time-dependent density functional theory. 

A reformulation of NE steady state quantum statistical mechanics has been
proposed by Hershfield in Ref.~[\onlinecite{Hershfield:1993}].
An explicit expression for the NE density matrix was derived for a system at a unique temperature
in the presence of an applied bias between two electrodes. A scheme upon which one can build 
non-perturbative calculations was also provided.
Such an approach has been successfully applied in numerical calculations
of quantum electron transport \cite{Schiller:1995,Schiller:1998,Han:2006,Han:2007,
Han:2007b,Han:2010,Han:2010b,Dutt:2011,Han:2012}.
The universal aspects of NE currents in a quantum dot has also been explored by Doyon et al.
in a somewhat different, but related, approach \cite{Doyon:2006}. Another approach to 
calculate the asymptotic form of operators in NE quantum systems is given 
in Ref.~[\onlinecite{Gelin:2009}]. However, in these approaches, only charge current
was considered. The whole system is at a unique temperature and cannot
support any other energy/heat transport processes, happening in parallel with quantum
charge transport.

\HN{Furthermore, a generalisation of Hershfield scheme to the full time dependent problem 
with arbitrary initial conditions is provided in Ref.~[\onlinecite{Hyldgaard:2012}].
In this work, the author uses the formal scattering approach of Lippmann and Schwinger
\cite{Lippman:1950} to construct the  time-dependent NE density matrix. 
The construction of such a NE density matrix  should also be valid for leads at 
different temperatures, although this point was not explicitly addressed in Ref.[\onlinecite{Hyldgaard:2012}]. 
One of the important outputs of this work is that the time-dependent NE density matrix
is variational and therefore one can set up an efficient single-particle evaluation 
scheme for the steady-state Hershfield form \cite{note:rev1,note:rev2}.

However, the connection between the NESOM and the previous NE density matrix scheme
was overlooked by the author of Refs.[\onlinecite{Hershfield:1993,Hyldgaard:2012}] .
We address such a connection (for the steady state) in an explicitly 
and rigorous manner in the present paper.}

We suggest an extension of the
approach originally developed by Hershfield, to more general NE conditions: the presence 
of both a temperature and chemical potential gradients between two electrodes connected 
to the quantum open system. 
We show how to construct a NE density matrix when the two
reservoirs are at two different temperatures and at two different chemical potentials.
For that, we use some concepts developed for asymptotic steady-state operators 
in Ref.~[\onlinecite{Bernard:2013,Gelin:2009,Doyon:2006,Fujii:2007}], along the lines of 
the original work of Hershfield \cite{Hershfield:1993}. 
We obtain the generalised Gibbs-like expression for the corresponding NE density matrix.
The extra NE terms (extra from an equilibrium grand-canonical density matrix) characterise 
the entropy production in the open system and are related to the fluxes of particle and heat.
We also show that the generalised NE density matrix for the steady-state is fully 
compatible with the NESOM of Zubarev \cite{Fujii:2007}.

The paper is organised as follows.
In Sec.~\ref{sec:NEdensmat}, we show how the NE density matrix can be expressed in
terms of asymptotic scattering operators. We discuss in details the choice
of the initial conditions and partition of the system. We also provide a central
result for an iterative expansion of asymptotic steady-state operators.
In Sec.~\ref{sec:Deltamu_oneT}, we use this result to rederive the original expression
of NE density matrix for a system at a unique temperature. 
We postulate a generalisation of the NE density matrix for conditions including 
both heat and charge currents in Sec.~\ref{sec:Deltamu_deltaT}.
A rigorous proof of the equivalent between our generalised NE density matrix
and the more general McLennan-Zubarev NE statistical operator is given
in Sec.~\ref{sec:discuss}. 
An explicitly example of the calculation of the NE density matrix for a model system
is given in Sec.~\ref{sec:example}.
Conclusion are presented in Sec.~\ref{sec:ccl}. 
Some central mathematical expressions used to derive our
results are provided in the Appendices.

\section{Non-equilibrium density matrix from asymptotic scattering techniques}
\label{sec:NEdensmat}

\subsection{Generalities}
\label{sec:ave_opX}

The average of an arbitrary operator $X$ is given by
\begin{equation}
\label{eq:aveX}
\begin{split}
\langle X(t) \rangle = {\rm Tr}[\rho_0 X(t)] = {\rm Tr}[\rho(t) X] ,
\end{split}
\end{equation}
where the density matrix $\rho$, solution of the Liouville equation, is
given by
\begin{equation}
\label{eq:rhooft}
\rho(t) = e^{-iH(t-t_0)} \rho_0 e^{iH(t-t_0)} .
\end{equation}
Here $H$ is the total Hamiltonian of the system and $\rho_0$ is the initial
density matrix at time $t_0$. 
The trace in Eq.~(\ref{eq:aveX}) is taken over the appropriate
degree of freedom characterising the entire system.

Eq.~(\ref{eq:aveX}) can be re-arrange by using the property of cyclic 
permutation of the trace and the fact that the total Hamiltonian can be split into 
two parts:
$H=H_0+W$, with a reference Hamiltonian $H_0$ (for example an unperturbed Hamiltonian),
and a perturbation $W$. 
We have, using $u=t-t_0$, the following expression:
\begin{equation}
\label{eq:aveX_bis}
\begin{split}
\langle X(t) \rangle  	& = {\rm Tr}[\rho_0 e^{i H u} X e^{- i H u}] \\
			& = {\rm Tr}[e^{- i H u}\rho_0 e^{i H u} X] \\
			& = {\rm Tr}[\rho_0 e^{i H_0 u}  e^{- i H_0 u}  e^{i H u} X e^{- i H u}] \\
			& = {\rm Tr}[\bar S(\tau) \rho_0  \bar S^{-1}(\tau) X] ,
\end{split}
\end{equation}
where we use the fact that $\rho_0$ commutes with $H_0$ and introduce the notation
$\tau=-u$ and $\bar S(\tau)=e^{i H \tau} e^{- i H_0 \tau}$ [\onlinecite{note:rev3}].

\subsection{Set-up and initial conditions}
\label{sec:setup}

We consider a (finite size) central region $C$, connected two electrodes (left $L$ 
and right $R$) acting as thermal and particle reservoirs. These electrodes
are described within the thermodynamics limits, i.e. they are macroscopic (semi infinite).
Initially they are at their own equilibrium, characterized by two temperatures 
$T_L$ and $T_R$, and by two chemical potentials $\mu_L$ and $\mu_R$. 
Furthermore, we ignore the interaction between particles in the electrodes, although 
the central region $C$ may contain such kind of interaction.

We are interested in steady state regime, and therefore we take the initial state 
of the system to be in the far remote past $t_0\rightarrow -\infty$. The system 
is then characterised by an Hamiltonian $H_0$.
After all parts of the system are ``connected'' and after some time elapses, 
the full system is considered to reach a NE steady state. 
The system is then characterised (at time $t$)
by an total Hamiltonian $H = H_0 + W$.
This time $t$ is considered to be the ``now'' time (we might take it to be $t=0$ 
in the following, but only for convenience) and $t$ is far enough from $t_0$ 
so that all the interactions act fully on the system.

The questions related to the possibility of reaching a NE steady-state
have been addressed in Refs.~[\onlinecite{Ruelle:2000,Tasaki:2003,Frohlich:2003,Tasaki:2006,
Maes:2010,Tasaki:2011,Moldoveanu:2011,Cornean:2014}].
It is also been argued that a system will always reach a steady-state
if it is a (or if it is connected to another) system in the thermodynamic 
limit regardless the presence (or absence) of adiabatic switching of
the interactions \cite{Ojima:1989,Cornean:2008,Cornean:2014}.

We are now facing different possible choices to perform the separation
of the full Hamiltonian into $H_0$ and $W$.
We know that the full system is described by
\begin{equation}
\label{eq:def_H}
\begin{split}
H = \sum_{\alpha=L,R} \left( H_\alpha + V_{\alpha C} + V_{C\alpha} \right) + H^0_C + V_C^{\rm int} , 
\end{split}
\end{equation}
where $H_\alpha, H^0_C$ are the non-interacting Hamiltonians of the $\alpha=L,R$ electrodes
and of the central region $C$ respectively. The interaction between particles
in region $C$ is given by $V_C^{\rm int}$ and the coupling between region $C$ and the
$\alpha$ electrode is given by $V_{\alpha C}$.
We also consider that all the non-interacting Hamiltonians, $H_L,H^0_C,H_R$,
commute with each other and with the occupation numbers $N_\beta$ ($\beta=L,C,R$).
The commutators between the interaction part 
$V= V_C^{\rm int} + \sum_\alpha \left( V_{\alpha C} + V_{C\alpha} \right)$ and $N_\beta$ does not
vanish, i.e. $[V,N_\beta]\ne 0$.

There are basically two families of approaches: the partitioning and the
partition-free schemes.
In the latter \cite{Cini:1980,Stefanucci:2004b},
the three $L,C,R$ regions are initially connected and all at equilibrium,
i.e. initially, there is one single $T$ and one single $\mu^{\rm eq}$.
The applied bias between the electrodes is then introduced under the form of an 
external potential. The interaction between particles in the central region $C$ 
could be introduced in either the initial Hamiltonian $H_0$ or in the ``coupling'' term $W$.
Such an approach has been successfully applied for studying quantum electron transport
in systems at a single temperature \cite{Stefanucci:2004b,Stefanucci:2004a}. 
However, it does not not seems particularly well adapted for the study of both energy/heat 
and charge transport. The introduction of a temperature gradient between the electrodes 
in the partition-free method
appears difficult to perform, especially in the form of an external perturbation 
on the electron system.

Therefore, we focus here on the second kind of approaches based on partitioning 
the system.
Initially,  all regions $L,C,R$ are separated and are at their own equilibrium.
The macroscopic $L$ and $R$ regions are represented by a density matrix $\rho_{L,R}$ 
expressed in the grand canonical ensemble, 
with temperature $T_\alpha = 1/k\beta_\alpha$ and chemical potential $\mu_\alpha$.
The initial density matrix of the central region is assumed to take any arbitrary form
$\rho_C$ as this region is not in the thermodynamic limit.

There are still two ways to partition the system:
Case (a) we take the non-interacting Hamiltonian $H_0$ to be defined by $H_0=H_L + H_R$ 
and for the perturbation $W = H^0_C + V_C^{\rm int} + \sum_\alpha ( V_{\alpha C} + V_{C\alpha} )$.
Therefore, we have for the initial density matrix $\rho_0 = \rho_L \otimes \rho_R$.
Case (b) we take for $H_0$ all the non-interacting Hamiltonians of the three
regions
$H_0=H_L + H^0_C + H_R$ and $W = V_C^{\rm int} + \sum_\alpha ( V_{\alpha C} + V_{C\alpha} )$
contains only the coupling/interaction terms.
In this case, we have for the initial density matrix 
$\rho_0 = \rho_L \otimes \rho_C \otimes \rho_R$.
The question that arises now is the following: how should be defined the density matrix 
$\rho_C$ for the central region?

In the set-up we want to study, $\rho_C$ cannot be given by a canonical or a grand 
canonical ensemble. Otherwise it would imply the presence of the third reservoir 
characterised by its own temperature (and chemical potential).
Therefore, we need to define $\rho_C$ from a microcanonical ensemble.
The density matrix $\rho_C$ can be given either in a
pure state representation $\rho_C = \vert\Psi_C\rangle\langle\Psi_C\vert$, where
the ket $\vert\Psi_C\rangle$ represents any linear combination of the states
$\vert n\rangle$ of the central region $C$, or in a mixed state representation 
$\rho_C = \sum_n w_n \vert n\rangle\langle n\vert$, with probabilities $w_n$ such 
as $\sum_n w_n = 1$. The probabilities $w_n$ are not given by a Boltzmann or Gibbs
factor since they are not obtained from a canonical or grand canonical ensemble.

The choice of the initial preparation ($\{w_n\}$) of the central region 
seems quite arbitrary. However, we know that for the long time limit, when 
a system has reached a steady-state after an applied perturbation, the initial
correlations vanishes and a single steady-state is reached regardless to the
choice of initial conditions \cite{Ruelle:2000,Tasaki:2003,Frohlich:2003,Tasaki:2006,
Maes:2010,Tasaki:2011,Moldoveanu:2011,Cornean:2014}. 
Therefore any particular choice of the initial density matrix $\rho_C$ is not relevant.
This is however not the case for the transient regime \cite{Myohanen:2008,Myohanen:2009}.

Since we want to expand the results of Hershfield to heat and charge transport, as
well as the results of Ref.~[\onlinecite{Bernard:2013}] to the presence of a central 
region between the two electrodes, we are choosing similar initial conditions as in
Ref.~[\onlinecite{Bernard:2013}].
Hence we take the option (a) for the partitioning of the system, i.e.
$H_0=H_L + H_R$ and 
$W = H^0_C + V_C^{\rm int} + \sum_\alpha \left( V_{\alpha C} + V_{C\alpha} \right)$.

One should note that case (a) is related to case (b) when one takes, for initial 
condition for the central region, a density matrix $\rho_C$ with zero matrix
elements (diagonal matrix for a mixed state representation). In other words,
we consider that initially the central region is empty of electrons. The initial
matrix density $\rho_0$ has then a block of zeros in the subspace of the central 
region $C$ and the matrix elements of $\rho_\alpha$ in the subspaces of 
the $\alpha=L,R$ electrodes. 
In the asymptotic limit, the NE density matrix will have matrix elements spreading 
over all the three different subspaces.
Since the NE density matrix is independent of the initial conditions in the
steady state \cite{Ruelle:2000,Tasaki:2003,Frohlich:2003,Tasaki:2006,
Maes:2010,Tasaki:2011,Moldoveanu:2011,Cornean:2014}, one can take a convenient
choice for the initial conditions that makes the derivations more easily
tractable \cite{note1}.

Finally, the left and right electrodes are prepared in a Gibbs grand-canonical
ensemble with density matrices $\rho_\alpha$ ($\alpha=L,R$)
\begin{equation}
\label{eq:rho_LR}
\begin{split}
\rho_\alpha = \frac{1}{Z_\alpha} e^{-\beta_\alpha(H_\alpha-\mu_\alpha N_\alpha)} ,
\end{split}
\end{equation}
with $Z_\alpha={\rm Tr}[e^{-\beta_\alpha(H_\alpha-\mu_\alpha N_\alpha)}]$.
By definition, we have $[H_\alpha,H_\beta]=0$ and 
$[H_\alpha,N_\beta]=0$, hence
\begin{equation}
\label{eq:rho_0}
\begin{split}
\rho_0 = \rho_L \otimes \rho_R = \frac{1}{Z}
e^{-\sum_\alpha \beta_\alpha(H_\alpha-\mu_\alpha N_\alpha)}  ,
\end{split}
\end{equation}
where $Z={\rm Tr}[e^{-\sum_\alpha\beta_\alpha(H_\alpha-\mu_\alpha N_\alpha)}]$.

\subsection{Asymptotic steady-state NE density matrix}
\label{sec:NErho_def}

For the asymptotic steady state regime, we consider that the time different
$u=t-t_0$ goes to $\infty$ in Eq.~(\ref{eq:aveX_bis}), hence $\tau\rightarrow -\infty$.
This means that either the time $t$ is fixed and the initial time $t_0$
is the far remote past $t_0\rightarrow -\infty$, or $t_0$ is fixed and $t$ is the far remote
future.
In this case, the average for the NE asymptotic steady state is obtained from
\begin{equation}
\label{eq:NEaveX}
\begin{split}
\langle X \rangle^{\rm NE} & = \lim_{u\rightarrow +\infty} {\rm Tr}[e^{- i H u}\rho_0 e^{i H u} X] \\
			& = \lim_{\tau\rightarrow -\infty}  {\rm Tr}[\bar S(\tau) \rho_0  \bar S^{-1}(\tau) X] \\
			& = {\rm Tr}[\Omega^{(+)} \rho_0 \Omega^{(+)-1} X] \\
			& = {\rm Tr}[\rho^{\rm NE} X],
\end{split}
\end{equation}
where we use the definition of the M$\o$ller operator \cite{GellMann:1953,Akhiezer:1981,Bohm:1993,Baute:2001} 
\begin{equation}
\label{eq:Moller}
\begin{split}
\Omega^{(+)} = {\rm lim}_{\tau\rightarrow -\infty} e^{i H \tau} e^{-i H_0 \tau} .
\end{split}
\end{equation}

In Appendix \ref{app:moller}, we recall some definitions of the M$\o$ller operators
and prove one of their important properties: the intertwining relations.
Such a relation connects the non-interacting Hamiltonian
$H_0$ to the full Hamiltonian: $\Omega^{(+)}H_0=H\Omega^{(+)}$.

The NE density matrix $\rho^{\rm NE}$ in the steady
state is obtained as \cite{Fujii:2007,Bernard:2013}:
\begin{equation}
\label{eq:rho_NE}
\begin{split}
\rho^{\rm NE} & = \Omega^{(+)} \rho_0 \Omega^{(+)-1} \\
& = \frac{1}{Z} e^{-\beta_L ( H_L^+ - \mu_L N_L^+ ) -\beta_R ( H_R^+ - \mu_R N_R^+ ) } ,
\end{split}
\end{equation}
where the asymptotic operator $X^+$ is defined as $X^+ = \Omega^{(+)} X \Omega^{(+)-1} $ 
for any operator $X$.

Eq.~(\ref{eq:rho_NE}) is the starting point for deriving the NE density matrix in the
form given by Hershfield and for providing a generalisation to the cases including 
temperature gradients ($\beta_L\ne\beta_R$) and applied biases ($\mu_L\ne\mu_R$).
For completing our derivations, we use an important identity:
\begin{equation}
\label{eq:HH0}
\begin{split}
H_L^+ + H_R^+ & = \Omega^{(+)} (H_L+H_R) \Omega^{(+)-1} = \Omega^{(+)} H_0 \Omega^{(+)-1} \\
& = H \Omega^{(+)} \Omega^{(+)-1} = H .
\end{split}
\end{equation}

\section{Charge current at a unique temperature}
\label{sec:Deltamu_oneT}

For a system at a unique temperature ($\beta_L=\beta_R$) and with an applied bias  ($\mu_L\ne\mu_R$),
the NE density matrix, given in Eq.~(\ref{eq:rho_NE}), is rewritten as:
\begin{equation}
\label{eq:rho_NE_Hershfield}
\begin{split}
\rho^{\rm NE} 	&= e^{-\beta ( H_L^+ + H_R^+ - \mu_L N_L^+ - \mu_R N_R^+ ) } / Z \\
		&= \frac{1}{Z} e^{-\beta ( H - \Upsilon ) } ,
\end{split}
\end{equation}
where 
\begin{equation}
\label{eq:rho_NE_Yoperators}
\begin{split}
& \Upsilon = \mu_L N_L^+ + \mu_R N_R^+ = \Omega^{(+)} Y_0 \Omega^{(+)-1} , \\
& Y_0=  \mu_L N_L + \mu_R N_R .
\end{split}
\end{equation}

Eq.~(\ref{eq:rho_NE_Hershfield}) has just the same form as the NE density
matrix developed by Hershfield in [\onlinecite{Hershfield:1993}]. 
This result suggests that a series expansion of the asymptotic operator
$\Upsilon= \Omega^{(+)} Y_0 \Omega^{(+)-1}$
can be obtained following the prescriptions given in the original paper of
Hershfield [\onlinecite{Hershfield:1993}]. Hence the $\Upsilon$ operator
in Eq.~(\ref{eq:rho_NE_Hershfield}) and Hershfield $Y$ operator can be 
determined from the same iterative scheme. 
Therefore we have $\Upsilon = \sum_n \Upsilon_{n,I}$ with
$\Upsilon_{n,I}(t)$ following the iterative relation 
\begin{equation}
\label{eq:Yope_iter}
\partial_t \Upsilon_{n+1,I}(t) = -i [ \tilde{W}_I(t), \Upsilon_{n,I}(t) ] \ ,
\end{equation}
where the operators are given in the interaction representation, $X_I(t)=  e^{iH_0 t} X e^{-iH_0 t}$ 
($\tilde{W}_I$ includes the adiabatic factor $\tilde{W}_I(t) = e^{-\eta\vert t\vert} e^{iH_0 t} W e^{-iH_0 t}$ ) 
and with the initial value $\Upsilon_{0,I} = Y_0$
since $Y_0$ commutes with $H_0$ \cite{note3}. 

\HN{It should be noted that the construction of the NE density matrix for 
a system at unique temperature can also be found in Ref.~[\onlinecite{Hyldgaard:2012}]. This
paper provides a simpler explicit, but more formal, construction of the general
time-dependent NE density matrix by using scattering theory and the full time
evolution operator. The steady-state properties are recovered as an asymptotic limit.}

Finally, we can check an important property of the $\Upsilon$ operator.
Since $\Upsilon$ is a linear superposition of the operators $N_\alpha^+$ ($\alpha=L,R$),
we have
\begin{equation}
\label{eq:N+commutatorH}
\begin{split}
[N_\alpha^+, H] & = \Omega^{(+)} N_\alpha \Omega^{(+)-1} H - H \Omega^{(+)} N_\alpha \Omega^{(+)-1} \\
& = \Omega^{(+)} N_\alpha H_0 \Omega^{(+)-1} - \Omega^{(+)} H_0 N_\alpha \Omega^{(+)-1} \\
& = \Omega^{(+)} [ N_\alpha , H_0 ]  \Omega^{(+)-1} = 0 ,
\end{split}
\end{equation}
as by definition $N_\alpha$ commutes with the non-interacting Hamiltonian $H_0$.
Therefore the operator $\Upsilon(t)=\Upsilon$ is a constant of motion (a conserved quantity), 
with respect to the total 
Hamiltonian $H$. And the NE density matrix $\rho^{\rm NE}$, given by Eq.~(\ref{eq:rho_NE_Hershfield}),
is indeed a time-independent density matrix, as expected for the steady state \cite{note:rev4}.

\section{Simultaneous heat and charge currents}
\label{sec:Deltamu_deltaT}

In the presence of both a temperature gradient and a chemical potential gradient ($\beta_L\ne\beta_R$, $\mu_L\ne\mu_R$),
there is a simultaneous flow of energy/heat and charge between the two electrodes, through the central
region $C$.

We reformulate the general expression of the NE density matrix Eq.~(\ref{eq:rho_NE}) by
introducing first an average temperature \cite{Bernard:2013}, 
via an average $\bar\beta$ defined by $\bar\beta = (\beta_L+\beta_R)/2$.
Hence, the exponent in Eq.~(\ref{eq:rho_NE}) becomes:
\begin{equation}
\label{eq:exponent}
\begin{split}
& -\beta_L ( H_L^+ - \mu_L N_L^+ ) -\beta_R ( H_R^+ - \mu_R N_R^+ ) \\
& = - \bar\beta (H_L^+ + H_R^+) +(\bar\beta-\beta_L)H_L^+ + (\bar\beta-\beta_R)H_R^+ + \bar\beta Y^Q \\
& = - \bar\beta (H_L^+ + H_R^+) - \bar\beta Y^E + \bar\beta Y^Q  , 
\end{split}
\end{equation}
where 
\begin{equation}
\label{eq:YQE}
\begin{split}
\bar\beta Y^Q & =(\beta_L\mu_L N_L^+ + \beta_R\mu_R N_R^+)   ,\\
\bar\beta Y^E & =(\beta_L-\beta_R)\frac{1}{2}(H_L^+ - H_R^+) .
\end{split}
\end{equation}

The NE density matrix can be rewritten as
follows
\begin{equation}
\label{eq:rho_NE_genHershfield}
\begin{split}
\rho^{\rm NE} 	= \frac{1}{Z} e^{-\bar\beta ( H - Y^Q + Y^E ) }  .
\end{split}
\end{equation}
Note that in Eq.~(\ref{eq:rho_NE_genHershfield}), the generalised Gibbs-like form
of the NE density matrix is given with an effective temperature $\bar T$ defined 
from $\bar\beta$. This temperature is different from the temperature of the left 
or right electrodes $T_{L,R}$ since $\bar T = 1/k_B\bar\beta = 2 T_L T_R / (T_L + T_R)$.
 
The two quantities $Y^Q$ and $Y^E$ follow the same formal expression, i.e. 
$Y^x = c_L X_L^+ + c_R X_R^+ = \Omega^{(+)} (c_L X_L + c_R X_R)  \Omega^{(+)-1}$,
with $X_\alpha=N_\alpha$ ($H_\alpha$) for $Y^Q$ ($Y^E$ respectively).
Furthermore, Eq.~(\ref{eq:rho_NE_genHershfield}) has also the same formal structure 
of a generalised Gibbs ensemble as originally obtained by Hershfield.

We then suggest that the two quantities $Y^{Q,E}$ can be obtained
from the same formal iterative scheme:
\begin{equation}
\label{eq:iteration_QE}
\begin{split}
& Y^{Q,E} = \sum_n Y^{Q,E}_{n,I} , \\
& \partial_t Y^{Q,E}_{n+1,I}(t) = -i [ \tilde{W}_I(t), Y^{Q,E}_{n,I}(t) ]
\end{split}
\end{equation}
with the initial values  \cite{note4}
\begin{equation}
\label{eq:iteration_QE_initcond}
\begin{split}
Y^Q_{0,I} & = Y^Q_0 = a^Q_L N_L + a^Q_R N_R , \\
Y^E_{0,I} & = Y^E_0 = a^E (H_L - H_R ) ,  
\end{split}
\end{equation}
and
\begin{equation}
\label{eq:Y_QE_coef}
\begin{split}
& a^Q_\alpha= \frac{ 2 \beta_\alpha\mu_\alpha }{ \beta_L+\beta_R } \\
& a^E = \frac{ \beta_L-\beta_R }{ \beta_L+\beta_R } .
\end{split}
\end{equation}

Eq.~(\ref{eq:rho_NE_genHershfield}) and the iterative scheme, 
Eqs.~(\ref{eq:iteration_QE},\ref{eq:iteration_QE_initcond}) for $Y^{Q,E}$, 
are the main results of the paper.
We prove exact their formal equivalence with the McLennan-Zubarev form
of the NE density in the following section.
We also provide a concrete example for deriving the expression of the NE 
density matrix for a model system in Sec. \ref{sec:example}. 

The equations Eqs.~(\ref{eq:rho_NE_genHershfield},\ref{eq:iteration_QE},\ref{eq:iteration_QE_initcond})  
correspond to the most general expression of the steady-state NE density matrix 
in the presence of both heat and charge currents for a two-reservoir device.

As shown in the previous section, $Y^Q$ is a constant of motion since it commutes
with the total Hamiltonian $H$. It is easy to show that $Y^E$ is also a constant
of motion (a conserved quantity), 
since $[H_\alpha^+,H]=\Omega^{(+)} [ H_\alpha , H_0 ]  \Omega^{(+)-1} = 0$.

At equilibrium, $\beta_L=\beta_R$ hence $Y^E$ vanishes because $a^E = 0$. There is
a single chemical potential $\mu_L=\mu_R=\mu^{\rm eq}$
and $Y^Q = \mu^{\rm eq} (N_L^+ + N_R^+)$ [\onlinecite{note2}]. 
Hence one recovers the usual Gibbs form for
the equilibrium density matrix in a grand-canonical ensemble 
$\rho^{\rm eq} = e^{-\beta(H-\mu^{\rm eq} N)}/Z$, as expected.

It is important to note that $Y^Q$ exists because of the presence of the two different
chemical potentials $\mu_{L,R}$ and hence
it is related to the charge current. The quantity $Y^E$ exists because of the presence
of the temperature gradient ($\beta_L-\beta_R$) and hence is related to the energy/heat 
flow between the electrodes.
Indeed, following Ref.~[\onlinecite{Bernard:2013}], we have
\begin{equation}
\label{eq:Y_E_Doyon}
\begin{split}
 Y^E & = \frac{ \beta_L-\beta_R }{ \bar\beta} E^+ , \\
 E^+ & = \Omega^{(+)} E \Omega^{(+)-1} , \\ 
 E   & = \frac{1}{2}(H_L - H_R) .
\end{split}
\end{equation}
In the Heisenberg representation, the energy current operator is given
by $j_E(t)=\partial_t E(t) = i[H,E(t)]$ and $E(t) = \int_{-\infty}^{t} j_E(u) {\rm d}u $.
We assume that there is not current at $t_0=-\infty$ since the system is
decoupled, and the interaction Hamiltonian $W$ vanishes.
Furthermore, the operator $Y^Q$ can be rewritten as
\begin{equation}
\label{eq:Y_Q_Doyon}
\begin{split}
Y^Q & = \bar\mu N + \frac{ \Delta_\mu }{ \bar\beta } Q^+ , \\
\bar\mu & = ( \beta_L\mu_L + \beta_R\mu_R ) / ( \beta_L + \beta_R ), \\
\Delta_\mu & = \beta_L\mu_L - \beta_R\mu_R , 
\end{split}
\end{equation}
and
\begin{equation}
\label{eq:Q_Doyon}
\begin{split}
Q^+ & = \Omega^{(+)} Q \Omega^{(+)-1} , \\
Q &= \frac{1}{2}(N_L - N_R) .
\end{split}
\end{equation}
The charge current operator $j_Q(t)$ is related to the quantity $Q$ as
$j_Q(t)=\partial_t Q(t) = i[H,Q(t)]$ and $Q(t) = \int_{-\infty}^{t} j_Q(u) {\rm d}u$
(no current at $t_0=-\infty$ since the system is decoupled).
The two quantities $Y^{Q,E}$ are now clearly related to the charge and heat flows
induced by the NE conditions. They are also associated with the entropy production 
in the system \cite{Ness:2013,note:rev5}.
\HN{For the time dependent problem considered in Ref.~[\onlinecite{Hyldgaard:2012}]
for a system at a unique temperature, it was shown that the operator $Y$ describes
how the Gibbs free energy evolves as the interaction $W$ is adiabatically turned on.
}

\section{Connection with the McLennan-Zubarev NE statistical operator}
\label{sec:discuss}

In this section, we show the formal connection between of suggested generalisation of
the NE density matrix Eq.~(\ref{eq:rho_NE_genHershfield}) 
and the McLennan-Zubarev NE statistical operator.

In Ref.~[\onlinecite{Ness:2013}], we have shown that the original approach of Hershfield
provides a NE density matrix, see Eq.~(\ref{eq:rho_NE_Hershfield}), which
is a subset of the more general NE density matrix given by the McLennan-Zubarev method.

The generalised NE density matrix Eq.~(\ref{eq:rho_NE_genHershfield}) is 
fully compatible with the McLennan-Zubarev NE statistical operator method (NESOM).
Indeed, the quantities $E(t)$ and $Q(t)$, related to the
operators $Y^{E,Q}$ respectively, are expressed in terms of the time
integral $\int_{-\infty}^0 J_S(u) {\rm d}u$ entering the definition
of the the McLennan-Zubarev NE statistical operator \cite{Zubarev:1994,Tasaki:2006,Ness:2013}.
The latter is given by \cite{note_Sign}
\begin{equation}
\label{eq:rhoNESOM}
\begin{split}
\rho^{\rm NESO} = \frac{1}{Z}\
{\rm exp} \left\{ - \sum_\alpha \beta_\alpha \left( H_\alpha - \mu_\alpha N_\alpha \right) \right. \\  
\left. + \int_{-\infty}^0 {\rm d}s\ e^{\eta s} J_S(s) \right\} .
\end{split}
\end{equation}

The quantity $J_S(u)$ is called the non-systematic energy flows \cite{Tasaki:2006}
and is related to the entropy production rate of the system \cite{Ness:2013}.
It is given by 
\begin{equation}
\label{eq:Zub}
\begin{split}
J_S(s) & = \sum_\alpha \beta_\alpha J^q_\alpha(s) , \\
J^q_\alpha(s) & =  \frac{d}{du} ( H_\alpha(s) - \mu_\alpha N_\alpha(s) )  
\end{split}
\end{equation}
where all operators are given in the Heisenberg representation.

\HN{The concept of the non-systematic energy flows in the NESOM is also
consistent with the Gibbs free energy description given in Ref.~[\onlinecite{Hyldgaard:2012}]
(in this work, the variational for the NE density matrix and the corresponding
thermodynamic grand potential corresponds to an entropy maximization principle,
constrained by both the particle flow effects and the internal energy minimization).}

In Appendix \ref{app:lemma}, we derive a lemma which shows how the
time integral of an operator in the Heisenberg representation can be
expanded into a series of operators, in the interaction representation, 
involving commutators with the interaction Hamiltonian $W_I$.
Hence the integral $\int_{-\infty}^0 J_S(u) {\rm d}u$ entering the definition 
of $\rho^{\rm NESO}$ can be expanded in a series similar to that
obtained for the $Y^{E,Q}$ operators defining of the 
generalised NE density matrix.
 
We now proceed with the formal derivation of the connection between
Eq.~(\ref{eq:rho_NE_genHershfield}) and Eq.~(\ref{eq:rhoNESOM}).
For that we first note that
\begin{equation}
\label{eq:A}
\begin{split}
J_S(s)
& = i \sum_\alpha [ H , \beta_\alpha H_\alpha(s) - \beta_\alpha \mu_\alpha N_\alpha(s) ] \\
& = i \sum_\alpha e^{iHs} [W, \beta_\alpha H_\alpha - \beta_\alpha \mu_\alpha N_\alpha ] e^{-iHs} \ ,
\end{split}
\end{equation}
then we rewrite $\sum_\alpha \beta_\alpha H_\alpha$ as follows
\begin{equation}
\label{eq:B}
\begin{split}
\beta_L H_L + \beta_R H_R  = \bar\beta & (H_L + H_R + W ) \\
- \bar\beta W + & (\beta_L - \bar\beta) H_L + (\beta_R - \bar\beta) H_R \\
  = \bar\beta & ( H - W + Y^E_0 )
\end{split}
\end{equation}
with the help of Eq.~(\ref{eq:iteration_QE_initcond}).

Hence the NE statistical operator in Eq.~(\ref{eq:rhoNESOM}) can be reformulated
as 
\begin{equation}
\label{eq:rhoNESOM_2}
\begin{split}
\rho^{\rm NESO} = \frac{1}{Z}\
{\rm exp} - \bar\beta\left\{ H - W + Y^E_0 - Y^Q_0  \right. \\
\left.  -\int_{-\infty}^0 {\rm d}s\ e^{\eta s} J_S(s)/ \bar\beta \right\} .
\end{split}
\end{equation}

The integral of the non-systematic energy flows is obtained from different
contributions:
\begin{equation}
\label{eq:JS}
\int_{-\infty}^0 {\rm d}s\ e^{\eta s} J_S(s)/ \bar\beta = B^Q + B^E_{\rm tot} ,
\end{equation}
where
\begin{equation}
\label{eq:BQ}
\begin{split}
B^Q
& = \int_{-\infty}^0 {\rm d}s\ e^{\eta s} e^{iHs} (-i)[W, \sum_\alpha \beta_\alpha \mu_\alpha N_\alpha / \bar\beta] e^{-iHs} \\
& = \int_{-\infty}^0 {\rm d}s\ e^{\eta s} e^{iHs} \left( -i [ W, Y^Q_0 ] \right) e^{-iHs} .
\end{split}
\end{equation}
and
\begin{equation}
\label{eq:BEtot}
\begin{split}
B^E_{\rm tot} 
& = \int_{-\infty}^0 {\rm d}s\ e^{\eta s} e^{iHs} i[W, \sum_\alpha \beta_\alpha H_\alpha / \bar\beta] e^{-iHs} \\
& = \int_{-\infty}^0 {\rm d}s\ e^{\eta s} e^{iHs} \left( i [ W, H - W + Y^E_0 ] \right) e^{-iHs} ,
\end{split}
\end{equation}
using Eq.~(\ref{eq:B}).
The commutator in Eq.~(\ref{eq:BEtot}) contains three terms, the first is simply the time derivative of
the operator $W$ in the Heisenberg representation: $\partial_s W_H(s) = e^{iHs} i[H,W] e^{-iHs}$. Hence the time integral
(with the adiabatic factor) simply gives the value $-W_H(s=0)=W$. The second term in the commutator vanishes, while the
third term is:
\begin{equation}
\label{eq:BE}
\begin{split}
- B^E = \int_{-\infty}^0 {\rm d}s\ e^{\eta s} e^{iHs} \left( i [ W, Y^E_0 ] \right) e^{-iHs} .
\end{split}
\end{equation}
Hence 
\begin{equation}
\label{eq:JS_bis}
\int_{-\infty}^0 {\rm d}s\ e^{\eta s} J_S(s)/ \bar\beta = B^Q -W - B^E ,
\end{equation}
and the NE statistical operator $\rho^{\rm NESO}$ can be rewritten in a compact form
similar to Eq.~(\ref{eq:rho_NE_genHershfield}):
\begin{equation}
\label{eq:rhoNESOcompact}
\begin{split}
\rho^{\rm NESO} = \frac{1}{Z}\ e^{-\bar\beta ( H - \Upsilon^Q + \Upsilon^E ) }  ,
\end{split}
\end{equation}
with $\Upsilon^x=Y_0^x + B^x$ and $x\equiv Q,E$. We can now prove that the
quantities $\Upsilon^{Q,E}$ obey the same series iterative expansion as their
counterparts $Y^{Q,E}$ in Eq.~(\ref{eq:iteration_QE}). This is readily done
by using the Peletminskii lemma described in Appendix \ref{app:lemma}.
Indeed according to the lemma, the quantity
$B^x = \int_{-\infty}^0 {\rm d}s\ e^{\eta s} e^{iHs} \left( -i [ W, Y^x_0 ] \right) e^{-iHs}$ 
is strictly equal to
\begin{equation}
\label{eq:Bx}
\begin{split}
B^x = \int_{-\infty}^0 {\rm d}s\ e^{\eta s} e^{iH_0s} \left( -i [ W, Y^x_0 ] -i [W, B^x ] \right) e^{-iH_0s} \\
= -i \int_{-\infty}^0 {\rm d}s\ [ \tilde{W}_I(s), Y^x_0] -i \int_{-\infty}^0 {\rm d}s\ [ \tilde{W}_I(s) , B^x_I(s)] .
\end{split}
\end{equation}
The first commutator in the left-hand-side of Eq.~(\ref{eq:Bx}) is just the definition of the time derivative of
the quantity $Y^x_{1,I}(s)$, as $\partial_s Y^x_{1,I}(s) = -i [ \tilde{W}_I(s), Y^x_0]$. Hence the corresponding integral
simply gives $Y^x_{1,I}(s=0)-Y^x_{1,I}(s=-\infty)=Y^x_{1,I}(s=0)$, as we assume that $\tilde{W}_I(s)$ vanishes at $s=-\infty$.
Therefore we have,
\begin{equation}
\label{eq:Bx_bis}
\begin{split}
B^x = Y^x_{1,I}(s=0) -i \int_{-\infty}^0 {\rm d}s\ [ \tilde{W}_I(s) , B^x_I(s)] ,
\end{split}
\end{equation}
By inserting the definition of $B^x$ itself in the right-hand-side commutator of Eq.~(\ref{eq:Bx_bis}), we
obtain the series expansion 
\begin{equation}
\label{eq:Bx_ter}
\begin{split}
B^x = Y^x_{1,I}(0) + \int_{-\infty}^0 {\rm d}s\ [ -i \tilde{W}_I(s) , Y^x_{1,I}(s)] \\ 
+ \int_{-\infty}^0 {\rm d}s\ [ -i \tilde{W}_I(s) , \int_{-\infty}^s {\rm d}s_1 [ -i W_I(s_1), 
\int_{-\infty}^{s_1} {\rm d}s_2\  \\ 
[ -i \tilde{W}_I(s_2), Y^x_0] ] ] + \dots
\end{split}
\end{equation}
Hence we obtain the expected series expansion $\Upsilon^x = Y^x_0 + Y^x_{1,I} + Y^x_{2,I} + \dots$ where the different
terms of the series are given by the iterative scheme
$Y^x_{n+1,I} = \int_{-\infty}^0 {\rm d}s\ [ -i \tilde{W}_I(s), Y^x_{n,I}(s)]$, or equivalently by the differential equations
defined in  Eq.~(\ref{eq:iteration_QE}).

We have therefore proven in an accurate formal way that there is a one-to-one correspondence between our generalisation
of the NE density matrix postulated in Sec.~\ref{sec:Deltamu_deltaT} and the exact general form of the McLennan-Zubarev 
NE statistical operation Eq.~(\ref{eq:rhoNESOM}), henceforth proving ({\it a posteriori}) the validity of our results 
given in Sec.~\ref{sec:Deltamu_deltaT}.

\section{An example}
\label{sec:example}

We now show an example for the lowest order expansion of the NE density matrix by
considering a non-interacting system and a simple description for the central region $C$.

In the absence of interaction, the Hamiltonian for the central region $C$ is simply 
given by $  H^0_C   = \varepsilon_0 d^\dag d $
where $d^\dagger$ ($d$) creates (annihilates) an
electron in the level $\varepsilon_0$.
The non-interacting electrodes are also described by a quadratic Hamiltonian 
$\alpha=L,R$ with $H_\alpha   = \sum_{k\alpha} \varepsilon_{k\alpha} c^\dag_{k\alpha} c_{k\alpha} $
where ${k\alpha}$ is an appropriate composite index to label the free electrons of the $\alpha$ electrode.
The coupling between the central region and the electrodes is given via some hopping matrix elements $t_{k\alpha}$,
and we have 
$\sum_\alpha ( V_{C\alpha} + V_{\alpha C} ) = 
\sum_{k,\alpha} t_{k\alpha} \left( c^\dag_{k\alpha} d + d^\dag c_{k\alpha}  \right)$. We recall
that, by definition, we have $W = H^0_C + \sum_\alpha ( V_{C\alpha} + V_{\alpha C} )$, and that the
only non vanishing anti-commutators are $\{d,d^\dag\}=1$ and  
$\{c_{k\alpha},c^\dag_{p\beta}\} = \delta_{k} \delta_{\alpha\beta}$.

We now proceed to calculate the operators $Y^x$ ($x=Q,E$) from the iterative scheme developed in
Sec. \ref{sec:Deltamu_deltaT}.
The zero-Th order is given by the definition of the operators $Y_0^x$, i.e.
$Y^Q_0 = a^Q_L N_L + a^Q_R N_R$ and $Y^Q_0 = a^E_L H_L + a^E_R H_R$ from Eq.(\ref{eq:iteration_QE_initcond}).
Note that we introduced (for later convenience) a new notation for $Y^Q_0$ where
$a^E_L=a^E=-a^E_R$ from Eq.(\ref{eq:iteration_QE_initcond}).

The first order contribution $Y^x_{1,I}$ involves the calculation of the commutator
$[W,Y^x_0]$ which is built from three different kinds of commutators
$[d^\dag d, c^\dag_{k\alpha} c_{k\alpha}]$, 
$[c^\dag_{p\beta} d, c^\dag_{k\alpha} c_{k\alpha}]$, 
$[d^\dag c_{p\beta}, c^\dag_{k\alpha} c_{k\alpha}]$. 
We find
\begin{equation}
\label{eq:WYQN1}
\begin{split}
[W,Y^Q_0] & = \sum_\alpha a^Q_\alpha \sum_k t_{k\alpha} \left( d^\dag c_{k\alpha} - c^\dag_{k\alpha} d\right) \\
          & = i \sum_\alpha a^Q_\alpha j^Q_\alpha ,
\end{split}
\end{equation}
with the conventional definition of the charge current operator
$j^Q_\alpha = - i \sum_k t_{k\alpha} ( d^\dag c_{k\alpha} - c^\dag_{k\alpha} d ) $; and
\begin{equation}
\label{eq:WYEN1}
\begin{split}
[W,Y^E_0] = i \sum_\alpha a^E_\alpha j^E_\alpha ,
\end{split}
\end{equation}
with the definition of the energy current operator
$j^E_\alpha = - i \sum_k \varepsilon_{k\alpha} t_{k\alpha} ( d^\dag c_{k\alpha} - c^\dag_{k\alpha} d ) $
by analogy with the definition of the charge current.
Therefore, the first order contribution $Y^x_{1,I}$ is obtained from
\begin{equation}
\label{eq:X1I}
\begin{split}
Y^x_{1,I} = \sum_\alpha a^x_\alpha \int_{-\infty}^0 {\rm d}s\ e^{\eta s}  j^x_{\alpha,I}(s) ,
\end{split}
\end{equation}
with $j^x_{\alpha,I}(s)$ being the interaction representation of $j^x_\alpha$ ($x=Q,E$).
Such a result can also been obtained, in a more straightforward way, from the expression
of the NESO given in Eq.~(\ref{eq:rhoNESOM}). Indeed, the first order contribution
$Y^x_{1,I}$ is simply obtained from the integral of the non-systematic energy flow by
replacing the Heisenberg representation of $J_S(s)$ by its lowest order expansion in
the interaction representation.

Interestingly, one can introduce an advanced quantity $f^{\rm adv}(s)$ by defining
$f^{\rm adv}(s) = \theta(-s) e^{\eta s} f(s)$.
Hence the time integral in Eq.~(\ref{eq:X1I}) 
becomes the Fourier transform  $\int_{-\infty}^{\infty} {\rm d}s\ j^{x,{\rm adv}}_{\alpha,I}(s)$ of
$j^{x,{\rm adv}}_{\alpha,I}(s)$ evaluated at $\omega=0$.
The first order contributions  
\begin{equation}
\label{eq:X1I_bis}
\begin{split}
Y^x_{1,I} = \sum_\alpha a^x_\alpha j^{x,{\rm adv}}_{\alpha,I}(\omega=0)
\end{split}
\end{equation}
are then related to the static (d.c.) limit of the current operators.

The higher order contributions are more cumbersome to evaluate explicitly. 
For example, the second order contributions are given by
\begin{equation}
\label{eq:X2I}
\begin{split}
Y^x_{2,I} & = \int_{-\infty}^0 {\rm d}s\ e^{\eta s} [-i \tilde{W}_I(s), Y^x_{1,I}(s)] 
= \sum_\alpha a^x_\alpha \\
& \int {\rm d}s \int {\rm d}s_1\ e^{\eta s} e^{\eta s_1} e^{iH_0s}
[-i W, j^x_{\alpha,I}(s_1-s) ] e^{-iH_0s} .
\end{split}
\end{equation}
Their evaluation involves not only the calculation of the commutator between $W$ and $j^x_\alpha$ 
but also the series expansion of $j^x_{\alpha,I}$ in terms of $H_0$.

For any perturbation series expansion, the results given by a lowest order expansion
of the NE density matrix will always be different from the exact (fully resumed) results.
We provide, in Appendix \ref{app:errors}, a brief analysis of the errors introduced by
a finite series expansion of the $Y^{Q,E}$ operators.

One can also draw some analogies between our results and the results for the expression of the operator 
$Y$ (system at a unique $T$) given in  Ref.~[\onlinecite{Han:2006}].
This can be done by introducing the definition of the advanced Green's function
$g_0^{\rm adv}(\omega) = i \int {\rm d}s e^{\eta s} \theta(-s) e^{\pm iH_0s} e^{i\omega s}
= [\omega \pm H_0 - i\eta]^{-1} $.
Furthermore the central quantities, in the interaction representation of the current operators,
are $e^{iH_0s} d^\dag c_{k\alpha} e^{-iH_0s}$ and $e^{iH_0s} c^\dag_{k\alpha} d e^{-iH_0s}$.
This quantities can be re-expressed as follows
\begin{equation}
\label{eq:truc}
\begin{split}
e^{iH_0s} d^\dag c_{k\alpha} e^{-iH_0s} & = - d^\dag c_{k\alpha} ( 1 + e^{-i\varepsilon_{k\alpha}s} e^{-iH_0s} ) \\
e^{iH_0s} c^\dag_{k\alpha} d e^{-iH_0s} & = - c^\dag_{k\alpha} d ( 1 - e^{ i\varepsilon_{k\alpha}s} e^{-iH_0s} ) ,
\end{split}
\end{equation}
and the time integration of the corresponding time-dependent factors will lead to the
appearance of the Green's functions in the series expansion of the $Y^{Q,E}$ terms.
 
One should however note that, by definition, our results are formally different from the expression 
of the operator $Y$ given in Ref.~[\onlinecite{Han:2006}].
We are dealing with a general problem where
$\beta_L \ne \beta_R$ and $\mu_L \ne \mu_R$ and the possibility of an asymmetric potential
drop, i.e. our $\mu_\alpha$ are different from the symmetric case $\mu_L = V/2 = - \mu_R$.
Furthermore, our expressions will be different from the results of Ref.~[\onlinecite{Han:2006}] 
since we are using a different initial density matrix $\rho_0 = \rho_L \otimes \rho_R$. 
We do not consider that initial the central region is described by a canonical ensemble 
$\rho_C \ne e^{-\beta H^0_C}/Z$ as explained in detail in Sec. \ref{sec:setup}.
 
\HN{Finally, we briefly comment on possible extensions to systems where
the interaction is not only limited to the central region. We suggest
that the essential point is that the reservoirs are indeed described by 
an equilibrium density matrix, hence interaction may exist in them and 
throughout the entire system. 
However, when considering the iterative scheme to calculate the $Y^{E,Q}$ 
quantities, one can anticipate that each iteration will involve operators 
which get more and more spread all over the entire system (when interaction 
exist inside the leads). This point might then lead to strong computational
constraints in comparison to the cases where the interaction is present only
in the central region.}

\section{Conclusion}
\label{sec:ccl}

We have proposed how to expand the NE density matrix originally developed by Hershfield 
to the cases of simultaneous (steady state) current flows of heat and charge.
The stationary density matrix of an open system is written in the generalised Gibbs form 
$\rho^{\rm NE}=e^{-\bar\beta(H-Y^Q+Y^E)}/Z$, 
with the nonequilibrium ``correction terms'' $Y^{Q,E}$ being related to the charge and
energy currents imposed by the NE conditions. 
We have provided an explicit iterative scheme to calculate the $Y^{Q,E}$ operators
which is similar to the iterative scheme developed originally by Hershfield.

We have also proved in a rigorous way that our generalised  NE density matrix is
strictly equivalent to the McLennan-Zubarev form of the NE statistical operator,
validating {\it a posteriori} the correctness of our scheme.

The operator $\bar\beta(Y^E-Y^Q)$ is related to the entropy production of the NE quantum 
open system \cite{note:rev5}. 
It can be calculated in the absence and in the presence of interaction and gives 
information about the dissipation in the driven system.
We have provided an explicit example for the lowest order expansion of the NE density
matrix for a non-interacting model system. 

The generalized scheme to calculate the NE density matrix that we have presented here 
can now serve as the basis for numerical calculations of both heat and charge transport 
using the numerical techniques developed on the original approach of Hershfield \cite{Han:2006,Han:2007,Han:2007b,Han:2010,Han:2010b,Dutt:2011,Han:2012}.
As clearly shown in Ref.[\onlinecite{Hyldgaard:2012}], the NE density matrix has
variational properties and hence can also be used to define a rigorous single-particle
scheme in the spirit of a density-functional-based theory (once the proper NE functionals
are properly set up) \cite{Hyldgaard:2012}.

The NE density matrix can also lead to more insight for the NE physical properties of 
quantum open systems and 
to the derivation of NE thermodynamical laws, such as NE fluctuation-dissipation relations 
\cite{Ness:2014b}, NE electron distribution function \cite{Ness:2014a} 
and NE charge susceptibility \cite{Ness:2012}.

Finally, we would like to make two general comments.
First, we briefly comment on the connections between the NE density matrix and
the more widely use NE Green's functions (GF) approaches.
 The GF are correlation functions whose thermodynamical averages are
formally identical to those calculated in Hershfield approach (which
we generalized in the present paper for the cases of two reservoirs at
different chemical potentials and temperatures). As we explained in
Ref.~[\onlinecite{Ness:2013}], both perturbation series used in the NE GF approach
and in the derivations of the equations for the $Y^{Q,E}$ operators
start from the same nonequilibrium series expansion. They are two
different ways of summing that series. For a non-interacting problem
for which the series can be resumed exactly, the NE GF and the
Hershfield $Y$ operator approach provide the same results \cite{Schiller:1995,Schiller:1998}.
For an interacting system, one must resort to approximations to re-sum partially the series, 
and therefore the two approaches are similar only when the same level of approximations are 
used \cite{Han:2006,Dash:2010,Dash:2011}.

Second, we want to point out that various authors have constructed a number of theoretical 
schemes for the
description of irreversible processes in NE systems.
In this paper, we have focussed on the approaches developed by Hershfield and Zubarev.
Other schemes have been developed by Peletminskii {\it et al.} to find
expressions for the NE density matrix 
\cite{Peletminskii:1971a,Peletminskii:1971b,Peletminskii:1972a,Peletminskii:1972b}.
A critical study of the equivalence between the two kinds of methods can be found
in Ref.~[\onlinecite{Zubarev:1971}]. 
It is interesting to note that in the Peletminskii approaches, the solution 
is related to a series expansion of the density matrix, as obtained from
a perturbation expansion of the solution of the Liouville equation for the density matrix 
(and with the 
appropriate sources term that ensure the irreversible nature of the time evolution) \cite{Note:5}
In the other approaches, the NE density matrix is also given by a series expansion
but rather in the form of a linked-cluster-like expansion \cite{Mahan:1990}, i.e. 
the corresponding series expansion enters the argument of an exponential functional.
In principle, if all the resummations are performed exactly the two kinds of approach
are also equivalent.

\begin{acknowledgments}
The author thanks Benjamin Doyon and Lev Kantorovich 
for insightful comments and useful discussions.
\end{acknowledgments}

\appendix

\section{The M$\o$ller operators for scattering}
\label{app:moller}

By definition \cite{GellMann:1953,Akhiezer:1981,Bohm:1993,Baute:2001},
the M$\o$ller operators are given for two asymptotic limits:
\begin{equation}
\label{eq:moller_limit}
\begin{split}
\Omega^{(\pm)} = {\rm lim}_{t\rightarrow \mp\infty} e^{i H t} e^{-i H_0 t} .
\end{split}
\end{equation}
corresponding to retarded or advanced evolution of the system.
Alternatively, the M$\o$ller operators can be also expressed in an integral form
\cite{Akhiezer:1981,Bohm:1993,Baute:2001}:
\begin{equation}
\label{eq:moller_integral}
\begin{split}
\Omega^{(\pm)} = {\rm lim}_{\eta\rightarrow 0^+} (\mp\eta)
\int_{0}^{\mp\infty} {\rm d}u\ e^{\pm\eta u} e^{i H u} e^{-i H_0 u} .
\end{split}\end{equation}
They follow the intertwining property: $\Omega^{(\pm)}H_0=H\Omega^{(\pm)}$
which we now prove for $\Omega^{(+)}$.
We start by writing: 
\begin{equation}
\label{eq:moller_demoA}
\begin{split}
\Omega^{(+)} & = {\rm lim}_{\eta\rightarrow 0^+} 
\int_{-\infty}^0 {\rm d}\tau\ \eta e^{\eta\tau} e^{i H\tau} e^{-i H_0\tau} \\
& =  {\rm lim}_{\eta\rightarrow 0^+} 
\int_{-\infty}^0 {\rm d}\tau\ \eta e^{\eta\tau} \bar{S}(\tau) .
\end{split}
\end{equation}
It can be seen that $\bar{S}(\tau)$ obeys the following differential equation:
$\partial_\tau\bar{S}(\tau)= i \bar{S}(\tau) W_I(\tau)$.
Hence
\begin{equation}
\label{eq:moller_demoB}
\begin{split}
H\Omega^{(+)} & = {\rm lim}_{\eta\rightarrow 0^+}
\int_{-\infty}^0 {\rm d}\tau\ \eta e^{\eta\tau} e^{i H\tau} H e^{-i H_0\tau} \\
& =  \Omega^{(+)} H_0 + {\rm lim}_{\eta\rightarrow 0^+}
\int_{-\infty}^0 {\rm d}\tau\ \eta \bar{S}(\tau) \tilde{W}_I(\tau),
\end{split}
\end{equation}
where the adiabatic term has been included in $\tilde{W}_I(\tau)$.
The integral in Eq.~(\ref{eq:moller_demoB}) becomes
\begin{equation}
\label{eq:moller_demoC}
\begin{split}
\eta \int_{-\infty}^0 {\rm d}\tau \bar{S}(\tau) {W}_I(\tau)
= -i\eta \int_{-\infty}^0 {\rm d}\tau  \partial_\tau\bar{S}(\tau) \\
= -i\eta [\bar{S}(\tau) ]_{-\infty}^{0} = -i\eta (1-\bar{S}(-\infty) ),
\end{split}
\end{equation}
and vanishes in the limit $\eta\rightarrow 0^+$ since 
$\bar{S}(-\infty)=\bar{S}(-\infty)^\dag = \Omega^{(+)-1}$ is finite.

Hence the relation $H\Omega^{(+)} = \Omega^{(+)} H_0$
is proved.

\section{The Peletminskii Lemma}
\label{app:lemma}

In this section, we consider a useful Lemma given by Peletminskii in the Appendix
of Ref.~[\onlinecite{Peletminskii:1972a}]. We rederive the lemma below since we use 
a different sign convention and include an adiabatic factor $e^{\eta x}$.

The lemma provides, in an integral form, a connection between
the Heisenberg representation of an operator and the corresponding series
expansion of operators in the interaction representation.

Suppose that, for an arbitrary operator $A$, we define
\begin{equation}
\label{eq:pelet_def_B}
\begin{split}
B = \int_{-\infty}^{0} {\rm d}x\ e^{\eta x} e^{iHx} A e^{-iHx} =\int_{-\infty}^{0} {\rm d}x\ e^{\eta x} A_H(x) 
\end{split}
\end{equation}
where $A_H(x)$ is the Heisenberg representation of $A$ with respect to 
the total Hamiltonian $H=H_0+W$, and $\eta\rightarrow 0^+$.

Introducing, an intermediate quantity:
\begin{equation}
\label{eq:pelet_def_Abar}
\begin{split}
\bar A(x) = e^{-iH_0x} e^{iHx} A e^{-iHx} e^{iH_0x} , 
\end{split}
\end{equation}
we can see that $\bar A(x)$ follows the differential equation
$\partial_x \bar A(x) = i [W_I(-x),\bar A(x)]$ since 
the quantity $P(x) =  e^{-iH_0x} e^{iHx} = \bar{S}^{-1}(-x)$ obeys $\partial_x P(x) = i W_I(-x) P(x)$.
Hence
\begin{equation}
\label{eq:pelet_def_Abarx}
\begin{split}
\bar A(\tau) = A + i \int_{0}^{\tau} {\rm d}x\ [W_I(-x),\bar A(x)], 
\end{split}
\end{equation}
where $W_I(x)$ is the interaction representation of $W$: $W_I(x)= e^{iH_0x} W e^{-iH_0x}$.

By reversing the definition Eq.~(\ref{eq:pelet_def_Abar}) and using Eq.~(\ref{eq:pelet_def_Abarx}),
we find that
\begin{equation}
\label{eq:pelet_B_expand_1}
\begin{split}
B = & \int_{-\infty}^{0} {\rm d}x\ e^{\eta x} e^{iH_0x} A e^{-iH_0x} \\
+ i & \int_{-\infty}^{0} {\rm d}x\ e^{\eta x} \int_{0}^{x} {\rm d}y\ e^{iH_0x} [W_I(-y),\bar A(y)]  e^{-iH_0x} .
\end{split}
\end{equation}

Now, we follow two steps of calculation: (1) use the definition of $W_I(-y)$ and change the
variable $y$ into $v=x-y$, and (2) use the definition of $\bar A(x)$, to transform Eq.~(\ref{eq:pelet_B_expand_1})
into
\begin{equation}
\label{eq:pelet_B_expand_2}
\begin{split}
B = & \int_{-\infty}^{0} {\rm d}x\ e^{\eta x} e^{iH_0x} A e^{-iH_0x} \\
- i & \int_{-\infty}^{0} {\rm d}x\ e^{\eta x} \int_{x}^{0} {\rm d}v\ e^{iH_0v} [W,A_H(x-v)]  e^{-iH_0v} .
\end{split}
\end{equation}

Finally, by swapping the order of the integrals 
$\int_{-T}^{0} {\rm d}x \int_{x}^{0} {\rm d}v \rightarrow  \int_{-T}^{0} {\rm d}v \int_{-T}^{v} {\rm d}x$
(with $T\equiv\infty$),
and identifying
$\int_{-\infty}^{v} {\rm d}x\ e^{\eta x} e^{iH(x-v)} A e^{-iH(x-v)} = e^{\eta v} B$,
we obtain the following lemma:
\begin{equation}
\label{eq:pelet_final}
\begin{split}
B & = \int_{-\infty}^{0} {\rm d}x\ e^{\eta x} e^{iHx} A e^{-iHx} \\
  & = \int_{-\infty}^{0} {\rm d}x\  e^{\eta x} e^{iH_0x} \left( A - i [W,B] \right) e^{-iH_0x} . 
\end{split}
\end{equation}

Eq.~(\ref{eq:pelet_final}) connects, in an integral form, the Heisenberg representation
of $A$ with a series expansion of commutators ($[\dots[W,[W,A]\dots]$) in the 
interaction representation.
The lemma Eq.~(\ref{eq:pelet_final}) is central to our proof of the equivalence between
the NE density matrix $\rho^{\rm NE}$ and the McLennan-Zubarev NE statistical operator.

\section{Convergence of the expected values}
\label{app:errors}

One can perform the calculation of the NE density matrix by using only a finite number
of terms in the series expansion of the $Y^{Q,E}$ (in Sec. \ref{sec:example} we show only the two
first terms). 
We call the corresponding NE
density matrix $\rho^{\rm NE}_{(n)}$, it is obtained from the lowest $n$-terms in the
series expansion 
of the operators 
\begin{equation}
Y^{Q,E}_{(n)} = \sum_{i=0}^{n} Y^{Q,E}_{i,I},
\end{equation}
instead of the full series given by Eq.~(\ref{eq:iteration_QE})

The corresponding error induced the average of any operator $X$ is given by:
\begin{equation}
\label{eq:errorNEaveX}
\begin{split}
\delta\langle X \rangle^{\rm NE}_{(n)} 
& =  {\rm Tr}[\rho^{\rm NE} X] - {\rm Tr}[\rho^{\rm NE}_{(n)} X] \\
& =  {\rm Tr}[ ( \rho^{\rm NE} - \rho^{\rm NE}_{(n)} ) X],
\end{split}
\end{equation}
with $\rho^{\rm NE}_{(n)} = \exp (-\bar\beta ( H - Y^Q_{(n)} + Y^E_{(n)} ) ) / {Z_{(n)}} $ 
and the partition function $Z_{(n)}={\rm Tr}[\rho^{\rm NE}_{(n)}]$.

We can now proceed with an analysis in terms of the power of the interaction $W^n$.
Both partition functions 
$Z={\rm Tr}[\rho^{\rm NE}]$ and $Z_{(n)}$ contain all
orders of the interaction $\mathcal{O}(W^n)$ with $n=0,1,2,3,\dots$.
In order to get the leading term (lowest power of $W^n$) in $\delta\langle X \rangle^{\rm NE}_{(n)}$,
we can just consider the difference of the two exponentials in the NE
densities. At the lowest order, it is easily found that
\begin{equation}
\begin{split}
e^{-\bar\beta ( H - Y^Q + Y^E ) } - e^{-\bar\beta ( H - Y^Q_{(n)} + Y^E_{(n)} ) } \\
\sim
\sum_{i=n+1}^\infty  \bar\beta ( Y^Q_{i,I} - Y^E_{i,I} ) ,
\end{split}
\end{equation}
which gives a leading term in $\mathcal{O}(W^{n+1})$.

Therefore, for the calculations shows in Sec. \ref{sec:example}, if we consider only the
terms up to $n=1$, the error is (for the non-interacting case) in $t_{k\alpha}^2$. Such
a lowest order expansion is only expected to be valid in the limit of weak coupling
between the central region and the electrodes.


\begin{thebibliography}{52}
\bibitem{note:rev2}
It should be noted that, when the steady-state limit is well-defined, the rigorous DFT-based 
scheme developed in Ref.~[\onlinecite{Hyldgaard:2012}] differs from the approaches based on
TDDFT, and is distinctly different from the approximated scheme called ``NEGF-DFT''.
\bibitem{note1}
For any arbitrary initial density matrix $\rho_C$,
the asymptotic steady state density matrix 
$\rho^{\rm NE}$ is given by  $\rho^{\rm NE} = \rho_L^{(+)} \otimes \rho_C^{(+)} \otimes \rho_R^{(+)}$
where $\rho_\alpha = \Omega^{(+)} \rho_\alpha \Omega^{(+)-1}$. The matrix density
of the central region can be seen as a functional of the corresponding non-interacting
Hamiltonian $\rho_C=\rho_C(H^0_C)$.
For $\rho_C$ given in the microcanonical ensemble, we have
$\rho_C = Z_C^{-1} \sum_n \vert n\rangle \delta_\Delta(\epsilon_n-E_C) \langle n\vert$
with the eigenstates $H^0_C\vert n\rangle = \epsilon_n\vert n\rangle$.
The $\delta_\Delta$ function is the ``regularized'' delta function defined by
$\delta_\Delta(\epsilon_n-E_C) = 1$ for $E_C\le \epsilon_n \le E_C + \Delta$ and 0
otherwise, and $Z_C={\rm Tr}_C[\delta_\Delta(H_C^0-E_C)]$.
Using the properties of the Moller operators, it can be shown that the scattering
state $\vert n^{(+)}\rangle = \Omega^{(+)}\vert n\rangle$, constructed from
an eigenstate of the non-interacting Hamiltonian, is also an eigenstate of the
total Hamiltonian with the same eigenvalue: $H\vert n^{(+)}\rangle = \epsilon_n\vert n^{(+)}\rangle$.
Hence the density matrix $\rho_C$ transforms into another
microcanonical density matrix $\rho_C^{(+)}=\Omega^{(+)} \rho_C \Omega^{(+)-1}
= Z_C^{-1} \sum_n \vert n^{(+)}\rangle \delta_\Delta(\epsilon_n-E_C) \langle n^{(+)}\vert 
= Z_C^{-1} \delta_\Delta^C(H-E_C)$, where $\delta_\Delta^C$ implies a summation
over the states of the central region $C$ only. 
The microcanonical density matrix $\rho_C^{(+)}$ does not contain any information 
about the NE conditions, i.e. about the flows of particles and/or heat. 
Hence it does not bring any information about the NE entropy production in the system.
These are contained in the density matrix $\rho_L^{(+)} \otimes \rho_R^{(+)}$ which
is explicitly considered in this paper.
\bibitem{note3}
One can note that, with our definition of the total
Hamiltonian, $Y_0$ also commutes with $H^0_C$. Therefore the only parts left
in $\tilde{W}_I(t)$, in the first iteration of Eq.~(\ref{eq:Yope_iter}), are 
those corresponding to the interactions 
$V= V_C^{\rm int} + \sum_\alpha ( V_{\alpha C} + V_{C\alpha} )$.
\bibitem{note4}
Once more, the only parts left in $\tilde{W}_I$, in the commutator $[ \tilde{W}_I(t), Y^{Q,E}_{0,I}]$
of the first iteration in Eq.~(\ref{eq:iteration_QE}), are those corresponding to the interactions 
$V= V_C^{\rm int} + \sum_\alpha ( V_{\alpha C} + V_{C\alpha} )$. As, by definition, the operators
generating $Y^{Q,E}_{0,I}$ (i.e. $N_{L,R}$ for $Y^Q$ and $H_{L,R}$ for $Y^E$) 
commute with $H^0_C$.
\bibitem{note2}
By definition, we have $[ N_L + N_R , H_0 ]=0$, hence

$\Omega^{(+)} [ N_L + N_R , H_0 ]  \Omega^{(+)-1} = [N_L^+ + N_R^+ , H] = 0$.
The quantity $N_L^+ + N_R^+$ can be seen as the total number of electrons $N$ 
for the complete system, which is an invariant of the problem since it 
commutes with the total Hamiltonian \cite{Bernard:2013}.
\end{thebibliography}
\end{document}